\documentclass[useAMS,usenatbib]{mnras}
\usepackage{amsmath,bm}
\usepackage{amssymb}
\usepackage{textcomp}
\usepackage{graphicx}
\usepackage{color}
\usepackage{bm}
\usepackage{xspace}
\usepackage{upgreek}
\usepackage{nicefrac}
\usepackage[normalem]{ulem}
\usepackage[single=true]{acro}

\renewcommand{\upi}{\uppi}
\newcommand{\be}{\begin{equation}}
\newcommand{\ee}{\end{equation}}

\newcommand{\ba}{\begin{eqnarray}}
\newcommand{\ea}{\end{eqnarray}}

\usepackage{pgffor}
\makeatletter
\def\dif{\@ifnextchar[{\@with}{\@without}}
\def\@with[#1]#2{
  \ensuremath{\frac{\foreach \x in {#2}{\mathrm{d}\x\,}}{\foreach \x in {#1}{\mathrm{d}\x\,}}}
}
\def\@without#1{
  \ensuremath{%
    \ifx\hfuzz#1\hfuzz
    \mathrm{d}
    \else
    \foreach \x in {#1}{\mathrm{d}\x\,}
    \fi
    }
}
\makeatother

\DeclareAcronym{hd}{
  short = HD ,
  long  = hydrodynamic ,
  class = hydro ,
  first-style = default 
}

\DeclareAcronym{mhd}{
  short = MHD ,
  long  = \acifused{hd}{magneto-HD}{magnetohydrodynamic} ,
  class = hydro ,
  first-style = default 
}
\DeclareAcronym{rmhd}{
  short = RMHD ,
  long  = \acifused{mhd}{relativistic MHD}{\acifused{hd}{relativistic magneto HD}{relativistic magnetohydrodynamic}} ,
  class = hydro ,
  first-style = default
}
\DeclareAcronym{nt}{
  short = non-thermal ,
  long  = non-thermal ,
  first-style=empty 
}

\DeclareAcronym{he}{
  short = HE ,
  short-plural-form = HE ,
  long  = high-energy ,
  long-plural-form = high energies ,
  class = hydro ,
  first-style = default
}
\DeclareAcronym{vhe}{
  short = VHE ,
  short-plural-form = VHE ,
  long  = very-high-energy ,
  class = astro ,
  first-style = default 
}
\DeclareAcronym{1d}{
  short = 1D ,
  long  = one-dimensional ,
  class = hydro ,
  first-style = short
}
\DeclareAcronym{2d}{
  short = 2D ,
  long  = two-dimensional ,
  class = hydro ,
  first-style = short
}
\DeclareAcronym{3d}{
  short = 3D ,
  long  = three-dimensional ,
  class = hydro ,
  first-style = short
}

\DeclareAcronym{cd}{
  short = contact discontinuity ,
  short-plural-form = contact discontinuities ,
  long  = contact discontinuity ,
  long-plural-form = contact discontinuities ,
  class = hydro ,
  first-style = empty
}
\DeclareAcronym{ts}{
  short = termination shock ,
  long  = termination shock ,
  class = hydro ,
  first-style = empty
}
\DeclareAcronym{eos}{
  short = EOS ,
  long  =  equation of state ,
  long-plural-form = equations of state ,
  class = hydro ,
  first-style = default
}
\DeclareAcronym{pwn}{
  short = PWN,
  long  = pulsar wind nebula,
  short-plural = e,
  long-plural = e,
  class = astro,
  first-style = default
}
\DeclareAcronym{cwb}{
  short = CWB ,
  long  = colliding-wind binary,
  long-plural-form  = colliding-wind binaries,
  class = astro,
  first-style = default
}
\DeclareAcronym{grbs}{
  short = GRBS,
  long  = gamma-ray binary system ,
  first-style = default
}
\DeclareAcronym{muq}{
  short = \(\upmu\)Q,
  long  = microquasr ,
  first-style = default
}

\DeclareAcronym{cr}{
  short = CR ,
  long  = cosmic ray ,
  class = astro ,
  first-style = default
}
\DeclareAcronym{ns}{
  short = NS ,
  long  = neutron star ,
  class = astro ,
  first-style = default
}
\DeclareAcronym{ism}{
  short = ISM ,
  long  = interstellar medium ,
  class = astro ,
  first-style = default
}
\DeclareAcronym{cmb}{
  short = CMB ,
  long  = cosmic microwave background ,
  class = astro ,
  first-style = default
}
\DeclareAcronym{ic}{
  short = IC ,
  long  = inverse Compton ,
  class = astro ,
  first-style = default
}
\DeclareAcronym{kn}{
  short = KN ,
  long  = Klein-Nishina ,
  class = astro ,
  first-style = default
}
\DeclareAcronym{mec}{
  short = MEC ,
  long  = monochromatic emission coefficient ,
  class = astro ,
  first-style = default
}

\DeclareAcronym{psrb}{
  short = PSR~B1259 ,
  long  = \protect{PSR~B1259$-$63/LS2883} ,
  class = object ,
  first-style = default
}
\DeclareAcronym{j1018}{
  short = J1018 ,
  long  = \protect{1FGL~J1018.6$-$5856} ,
  class = object ,
  first-style = default
}
\DeclareAcronym{j2032}{
  short = J2032 ,
  long  = \protect{PSR~J2032$+$4127/MT91~213} ,
  class = object ,
  first-style = default
}
\DeclareAcronym{j0632}{
  short = J0632 ,
  long  = \protect{HESS~J0632$+$057} ,
  class = object ,
  first-style = default
}
\DeclareAcronym{ss433}{
  short = SS~433 ,
  long  = \protect{SS~433/W50} ,
  class = object ,
  first-style = default
}

\DeclareAcronym{fer}{
  short = {\it Fermi} LAT ,
  long  = {\it Fermi} Large Area Telescope ,
  class = object ,
  first-style = default
}

\newcommand{\fer}{\ac{fer}\xspace}

\newcommand{\magic}{MAGIC\xspace}
\newcommand{\hess}{H.E.S.S.\xspace}
\newcommand{\veritas}{{\itshape Veritas}\xspace}

\newcommand{\psrb}{\ac{psrb}\xspace}
\newcommand{\fglj}{\ac{j1018}\xspace}
\newcommand{\lmc}{LMC P3\xspace}
\newcommand{\ls}{{LS~5039}\xspace}
\newcommand{\lsi}{{LSI+61\({}^\circ\)303}\xspace}
\newcommand{\ssw}{\ac{ss433}\xspace}
\newcommand{\cyg}{Cygnus X-3\xspace}
\newcommand{\cygl}{Cygnus X-1\xspace}
\newcommand{\muq}{\ac{muq}\xspace}
\newcommand{\eC}{\(\upeta\)Carinae\xspace}

\newcommand{\IC}{\ac{ic}\xspace}

\newcommand{\CD}{\ac{cd}\xspace}

\newcommand{\NT}{\ac{nt}\xspace}

\newcommand{\cd}{\textsc{cd}}
\newcommand{\psr}{\textsc{psr}}

\newif\iflong
\longfalse

\title[Strongly magnetized pulsar wind in binary systems]{Modelling the interaction between relativistic and non-relativistic winds
in binary pulsar systems: strong magnetization of the pulsar wind}
\author[S. V. Bogovalov et al.]{S.V.Bogovalov$^1$, D.Khangulyan$^{2}$\thanks{E-mail:
d.khangulyan@rikkyo.ac.jp}, A.Koldoba$^{3}$,
G.V.Ustyugova$^{4}$,F.Aharonian$^{5,6,1}$ \\
    $^{1}$National Research Nuclear University (MEPhI), Kashirskoje shosse, 31, Moscow, Russia\\
    $^{2}$Rikkyo University, 3-34-1, Nishi-Ikebukuro, Toshima-ku, Tokyo 171-8501, Japan\\
    $^{3}$Moscow institute of physics and technology, Institutskiy per. 9, Dolgoprudny, Russia\\
    $^{4}$Keldysh Institute of Applied Mathematics RAN, Miusskaya sq. 4, Moscow, Russia\\
    $^{5}$Dublin Institute for Advanced Studies, School of Cosmic Physics, 31 Fitzwilliam Place, Dublin 2, Ireland\\   
    $^{6}$Max-Planck-Institut f\"ur Kernphysik, Saupfercheckweg 1, 69117 Heidelberg, Germany
}

\begin{document}

\date{}

\pagerange{\pageref{firstpage}--\pageref{lastpage}} \pubyear{2019}

\maketitle

\label{firstpage}

\begin{abstract}
  We present a numerical study of the properties of the flow produced by the collision of a magnetized anisotropic
  pulsar wind with the circumbinary environment. We focus on studying the impact of the high wind magnetization on the
  geometrical structure of the shocked flow. This work is an extension of our earlier studies that focused on a purely
  hydrodynamic interaction and weak wind magnetization. We consider the collision in the axisymmetric approximation,
  that is, the pulsar rotation axis is assumed to be oriented along the line between the pulsar and the optical star.
  The increase of the magnetization results in the expansion of the opening cone in which the shocked pulsar wind
  propagates.  This effect is explained in the frameworks of the conventional theory of collimation of magnetized
  winds. This finding has a direct implication for scenarios that involve Doppler boosting as the primary mechanism
  behind the GeV flares detected with the \acl{fer} from \acl{psrb}.
  The maximum enhancement of the apparent emission is determined by the ratio of \(4\upi\) to the
  solid in which the shocked pulsar wind propagates. Our simulations suggest that this enhancement factor is decreased
  by the impact of the magnetic field.
  \acresetall
\end{abstract}


\section{Introduction}
\label{sec:intro}

Several binary systems, e.g., \psrb, \ls, \fglj, \lsi, and \lmc emit bright \NT emission that spans across the entire
electromagnetic spectrum, from radio to gamma-ray energies \citep[see, e.g.,][and references
theirin]{2013A&ARv..21...64D}. These systems are detected with ground-based Cherenkov detectors, \hess, \magic, and
\veritas in the \ac{vhe} regime and with the \fer at \acp{he} \citep[see, e.g.,][and references
therein]{2016ApJ...829..105C,2019arXiv190810764C,2017AIPC.1792d0017B,2017ICRC...35..729M,2018IJMPD..2744010L}.  The spectral maximum typically appears at
MeV-GeV energies.
The emission detected from these systems shows a complex orbital
phase dependence that yet have not obtained a detailed explanation.

In analogy to binary systems bright in the X-ray band, these systems were lumped together into a new class of
\ac{grbs}. While two \acp{grbs}, \ac{psrb} and \ac{j2032}, harbor confirmed radio pulsars \citep{johnston92,2015MNRAS.451..581L}, the nature of the compact objects
in other \acp{grbs} is not constrained observationally. Because of spectral similarities, it is commonly assumed that
all \acp{grbs} consist of non-acreting pulsars orbiting luminous stars \citep{dubus06}. The interaction of the winds from the
pulsar and the optical companion presumably triggers particle acceleration and consequent \NT emission
\citep{maraschi,1994ApJ...433L..37T,motch}. We note here that some binary systems detected in \ac{he} or even in the
\ac{vhe} regime do not belong to the class of \ac{grbs}. The gamma-ray emission detected from Galactic \muq \cygl,
\cyg, and \ssw constitutes a small fraction, \(<10\%\), of the bolometric luminosity of the sources, and most-likely originates in the
jets \citep{2010MNRAS.404L..55D,2016A&A...596A..55Z,2018MNRAS.479.4399Z,2018Natur.562...82A,2019ApJ...872...25X} or even in the jet termination region
\citep{2015ApJ...807L...8B}.

Gamma-ray emission was also detected from \acp{cwb}, i.e., systems where two stars with powerful winds orbit each
other. \acp{cwb} are well-established sources of \NT emission \citep{1993ApJ...402..271E,2013A&A...558A..28D,2018NatAs...2..731H}.  In the
case of extended \ac{cwb}, e.g. WR~140 \citep[a WR$+$O binary system, see][]{2005mshe.work...49D,2005ApJ...623..447D},
radio observations allowed us to localize the acceleration site that appeared to be between the stars, where the wind
interaction is the most intense. One of the most extreme \ac{cwb}, \eC has been also detected as a gamma-ray source
with \fer \citep{2010ApJ...723..649A,2015A&A...577A.100R} and \hess \citep{2017ICRC...35..717L}. The detected \NT
emission from \ac{cwb} is consistent with models accounting for the particle acceleration at wind \acp{ts} and
 advection \citep{2014ApJ...789...87R}. When \ac{nt} particles move along the flow they lose energy due to
 synchrotron, \IC, and hadronic (in case of \NT protons) emission mechanisms. Adiabatic cooling may also play an
important role. As a result, \NT particles have a complex spacial-energy distribution. Convolution of this distribution
with target fields, which in turn also have a highly non-homogeneous anisotropic distribution, allows us to compute the
broadband emission spectra and lightcurves \citep{2014ApJ...789...87R}.

In the framework of the rotation-powered pulsar scenario, \acp{grbs} represent a version of \ac{cwb}, where one of the winds
is ultrarelativistic. The relativistic nature of the pulsar wind implies important \ac{hd} differences between
\acp{cwb} and \acp{grbs} \citep{2012A&A...544A..59B}. Since the regular \acp{cwb} appear to be significantly less efficient \NT emitters as
compared to \acp{grbs}, it is natural to attribute the dominant \NT activity in \acp{grbs} to processes taking
place in the pulsar wind. Thus, \acp{grbs} were considered as compactified \ac{pwn} located
in an environment with an unusually dense photon field \citep[see, e.g.,][]{1994ApJ...433L..37T,1999APh....10...31K,1999MNRAS.304..359C,2011A&A...535A..20B}.  However, detailed
\ac{hd} and \ac{mhd} simulations revealed critical differences as compared to (M)HD processes taking place around
isolated pulsars \citep{bakuk08,bakuk12}. {Certain similarities can be seen between binary pulsars and \acp{pwn} around bow-shock pulsars, which propagate through \ac{ism} with a super-sonic proper speed. This, however, concerns only the region close to the head of the bow shock. At large distances from the apex point, the structure of the flow starts to deviate considerably. In the case of binary pulsars the ram pressure of the stellar and pulsar winds decreases at large distance allowing an opened pulsar wind zone \citep{bakuk08}. For bow-shock nebulae, the ram pressure of \ac{ism} remains constant, which results in a closed pulsar wind zone.}

{Theoretical studies \citep{1993ApJ...402..271E,1996ApJ...459L..31W} and numerical simulations \citep{romero,bakuk08,bakuk12,okazaki,2013A&A...560A..79L} of colliding winds revealed} that the shocked flow propagates into a limited solid
angle on the binary system scale. The interaction of two winds results in re-acceleration of shocked pulsar
wind to relativistic speeds \citep{bakuk08,bakuk12,2012A&A...544A..59B,2013A&A...560A..79L}. The shocked pulsar wind material reaches 
relativistic bulk speeds on the characteristic binary separation scale. The relativistic motion may have a strong impact on the production
of \ac{nt} emission, and result in complicated time variability patterns
\citep{2012ApJ...753..127K,mitya14,2017A&A...598A..13D}. This might be the key factor that explains the complex orbital phase dependence of
the \NT emission seen from \acp{grbs}. In particular, Doppler boosting seems to be the most natural explanation for the
emission with a luminosity exceeding the pulsar spindown losses seen from \psrb with \fer during the GeV flare registered in 2017  \citep{2018ApJ...863...27J,2018RAA....18..152C}, which becomes even more significant with the increased estimate for the distance to the source \citep{2018MNRAS.479.4849M}.

The impact of Doppler boosting on the emission critically depends on two factors: the fluid bulk Lorentz factor and
the solid  angle, \(\Delta\Omega\), subtended by the outflow. Since the flow bulk acceleration proceeds because of a
\ac{hd} process, the increase of the bulk Lorentz factor is accomplished by the severe adiabatic cooling of the
plasma. Thus, the impact of the high bulk Lorentz factor on the emission intensity depends on the radiation mechanism,
and might be even counterintuitive: the synchrotron emissivity is expected to be severely weakened by the bulk
acceleration \citep{mitya14}. Thus, the solid angle into which the shocked pulsar wind propagates may have a dominant
impact on the emission intensity.

In the previous works 
\citep{bakuk08,bakuk12} we considered the collision of the stellar and pulsar winds in the specific cases of non-magnetized and
weakly magnetized pulsar winds. The calculations were performed for an axisymmetric \ac{2d} geometry. In this case, the
rotational axis of the pulsar is aligned with the line connecting the pulsar and the companion star. Re-acceleration of
the post-shock flow results in a rapid weakening of the magnetic field. Thus, there should be very little differences in
the interaction geometry in the cases of non-magnetized and weakly magnetized pulsar winds colliding with the stellar
wind. The geometry of the structure formed on the binary system scale depends on the ratio of the winds' ram pressures \citep[for
details, see][]{bakuk08}. The magnetic field weakening makes it to be dynamically unimportant allowing us to generalize our
conclusion to the case of the arbitrary orientation of the pulsar rotation axis.

We note, however, that the kinetic energy flux in the pulsar wind is expected to be highly anisotropic
\citep{2002MNRAS.336L..53B}, thus realistic calculations demand a \ac{3d} setup. Also \ac{3d} simulations are
needed to account for orbital motion in the system \citep{romero,okazaki,2015A&A...577A..89B}, although certain
conclusions also can be obtained with \ac{2d} approaches
\citep{2012A&A...544A..59B,2016MNRAS.456L..64B,2017MNRAS.471L.150B,2018MNRAS.479.1320B}.

The results obtained by \citet{bakuk12} show a qualitative difference between binary pulsars and \ac{pwn} created by
isolated pulsars. In the latter case, the magnetic field strength increases in the shocked pulsar wind until magnetic
stresses become dynamically important \citep{kennel84}. The simulations presented by \citet{bakuk12} concern however
only the case of a weakly magnetized pulsar wind, a property that is expected based on \ac{1d} and \ac{2d} \ac{mhd}
consideration of \ac{pwn}
\citep{kennel84,2002MNRAS.336L..53B,2003AstL...29..495K,2005MNRAS.358..705B,2004A&A...421.1063D}. A recent \ac{3d}
numerical simulation relaxes the requirement for a small magnetization in pulsar winds \citep{2014IJMPS..2860168P,2016JPlPh..82f6301O}. If
the magnetic field is strong at the pulsar wind \ac{ts} then it may appear to be dynamically important even in the case
of an accelerating bulk flow. The toroidal magnetic field, which is expected to be present in the pulsar wind, can lead to 
in the collimation of the outflow provided that the magnetic field is sufficiently strong
\citep{1999MNRAS.305..211B}. Studying the possibility of such a collimation is essential for the interpretation of
the complicated orbital phase dependence seen from \ac{grbs}. In this paper we extend the previous
studies \citep{bakuk08,bakuk12} to the case of a strongly magnetized pulsar wind. We limit our consideration to the
axisymmetric case.  In such a configuration, the preferred directions for the \ac{hd} and the magnetic collimation coincide, so
one may expect the strongest collimation effect.

As our benchmark case, we consider \ac{grbs} \psrb. The system consists of a \(\sim47.8\rm\,ms\) pulsar in an elongated
orbit, with eccentricity \(e = 0.87\), around a massive O-type companion
\citep{johnston92,psr1259param,2014MNRAS.437.3255S,2018MNRAS.479.4849M}.  During the last 25 years this system was
observed with different instruments in radio frequencies
\citep{1994MNRAS.268..430J,1996MNRAS.279.1026J,2005MNRAS.358.1069J,2011ApJ...732L..10M,2014MNRAS.437.3255S,2019arXiv190408429F},
optical wavelengths \citep{psr1259param,2016MNRAS.455.3674V}, in the X-ray band
\citep{1995ApJ...453..424K,2003PASJ...55..473M,2009ApJ...698..911U,2006MNRAS.367.1201C,2009MNRAS.397.2123C,2014MNRAS.439..432C,2015MNRAS.454.1358C,2011ApJ...730....2P,2015ApJ...806..192P},
GeV gamma rays \citep{2011ApJ...736L..11A,2015ApJ...811...68C, 2018RAA....18..152C,2018ApJ...863...27J}, and in the
\ac{vhe} regime \citep{2005A&A...442....1A,2009A&A...507..389A,2013A&A...551A..94H}. The detected emission consists of
several components: the optical and infrared emission from the massive companion and its decretion disk, pulsed radio
emission, and a variable broadband component that presumably is produced by \ac{nt} electrons accelerated at the
interface between the stellar and the pulsar wind \citep[see also][where one considered \ac{nt} particles accelerated at
the \ac{ts} caused by the impact of the Coriolis force in the same kind of scenario]{2013A&A...551A..17Z}.

\section{Model setup and simulation results}

In the simplest case, if one approximates both the stellar and the pulsar winds by isotropic non-magnetized outflows, the
flow formed at the winds' collision features axial symmetry \citep{bakuk08}.  To simulate the interaction of a
magnetized pulsar wind, the calculations should be performed in a \ac{3d} setup \citep[see, e.g.,][]{1993A&A...276..648B,2000ApJ...532..400W},
or one needs to adopt an additional assumption
that the pulsar rotation axis and the line connecting interacting stars are parallel. This assumption is not
critical in the case of a weak pulsar wind magnetization \citep{bakuk12}. However, in the case of a high wind
magnetization, this might be a rather artificial conjecture. We nevertheless adopt this configuration for our
simulations. The reason for that is the following: we aim to study the maximum possible collimation of the shocked pulsar
wind by the magnetic field.  This is achieved in the axisymmetric setup. {To perform the numerical simulation, we use a mathematical model and numerical algorithm, which are summarized in \citet{Koldoba2019}. }

{\ac{2d} axisymmetric solution can be significantly affected by the Coriolis force at some distance from the collision apex \citep{2011A&A...535A..20B}. The typical distance at which the pulsar trace bends significantly can be easily estimated in the limit \(\eta\ll1\). The shocked material follows in this case a spiral trajectory, which in polar coordinates, \((r,\phi)\), is defined as
  \be
  r(\phi) = D + v_{\rm w}\phi / \omega\,. 
  \ee
  Here \(\omega\) is the orbital angular velocity of the pulsar, \(v_{\rm w}\) is the stellar wind radial velocity, and \(D\) is the star separation distance. Close to the pulsar location, \(\phi=0\), the curvature of this trajectory is
  \be
  R_{\rm c} = r \frac{\left(1+\left(\frac{r'}r\right)^2\right)^{\nicefrac32}}{1 +2 \left(\frac{r'}r\right)^2-\left(\frac{r''}r\right)}\simeq D\frac{v_{\rm w}}{2v_{\rm orb}}\,,
  \ee
  where \(v_{\rm orb}\) is the orbital velocity of the pulsar. Thus, the dimensionless size of the region, where the trace bending is negligible is \(z\simeq R_{\rm c}/D\simeq v_{\rm w}/(2v_{\rm orb})\). For the conditions typical in \ac{grbs}, this corresponds to \(z\simeq5\), consistently with the size of the region where the impact of the orbital motion is small as obtained with numerical simulations \citep[see, e.g.,][]{2012A&A...544A..59B}. }

Under the approximation of axisymmetry, the flow is characterized by two parameters. The first parameter
is the ratio of the momentum rates of the pulsar and stellar winds.  Although the pulsar spindown losses are distributed between Poynting and kinetic energy, we still introduce this parameter in a form that accounts only for the kinetic energy, whose flux is assumed to be isotropic:
\begin{equation}
  \eta =  \frac{\dot{M}c\gamma_0}{\dot{M}_*v_{\rm w}} \ , \label{eq_eta}
\end{equation}
where \(\dot{M}\) and \(\dot{M}_*\) are the mass-loss rates of the pulsar and the optical star, respectively. At the
collision distance, and the pulsar wind is assumed to have
a bulk Lorentz factor of \(\gamma_0\). In the case of \ac{psrb}, one expects that the \(\eta\)-parameter is in the range
\(0.2\)~--~\(0.6\) if the pulsar interacts with the polar wind \citep{bakuk08}.  The impact of the \(\eta\)-parameter
has been extensively studied in our previous works \citep{bakuk08,bakuk12}. In what follows we adopt a fixed value for
this parameter, \(\eta=0.3\). In the case of a purely hydrodynamic interaction, one should expect the following
configuration: the opening angles of the relativistic \ac{ts} and the \ac{cd} are \(\sim50^\circ\) and \(\sim70^\circ\),
respectively \citep[{see Fig.11 in}][]{bakuk08}. The comparision with  this geometry, characterized by  the opening angles close to \(60^\circ\), should
allow us to study the impact of magnetic field.

If the \(\eta\)-parameter is small, $\eta \ll 1$, the location of the \CD (i.e., the surface that separates the shocked relativistic and non-relativistic winds)
on the line connecting the stars is defined as follows \citep{bakuk08}:
\begin{equation}
r_{\cd}\approx\sqrt{\eta}\,. 
\end{equation}
Here, $r_{\cd}$ is the distance to the pulsar expressed in units of the star separation
distance. 

Another parameter that determines the flow structure is the pulsar wind maximum magnetization, \(\sigma\). Since the magnetic field
in the pulsar wind is expected to be toroidal, the Poynting flux along the symmetry axis vanishes. We assume the total energy flux from the pulsar to be: 
\begin{equation}
\dif[\Omega]{L_\psr} = \frac{\gamma_0 \dot M c^2}{4\upi} \left(1+\sigma \sin^2(\theta)\right),
 \label{energy}
\end{equation}
where \(\theta\) is the polar angle and the solid angle element is \(\dif{\Omega}=\sin\theta\dif{\phi,\theta}\). This allows us to express the \(\sigma\)-parameter as
\begin{equation}
\sigma =  \frac32\left(\frac{L_\psr}{\gamma_0\dot{M}c^2}-1\right)\,.
\end{equation}
 Here, $L_\psr$ is the spindown luminosity of the pulsar. In the simulations, we keep the parameters of the stellar wind and the kinetic energy flux of the pulsar wind to be fixed. Thus, the increase of the sigma parameter results in a change of the pulsar spindown losses, \(L_\psr\). 

We are now interested in the range of \(\sigma=0.1\)~--~\(1\), which  extends the case of small magnetization, \(\sigma<0.1\), presented in \citet{bakuk12}, to the range of the magnetization consistent with \ac{3d} simulations \citep{2014IJMPS..2860168P,2016JPlPh..82f6301O}.

\begin{figure} 
\begin{center}
\includegraphics[width=100mm]{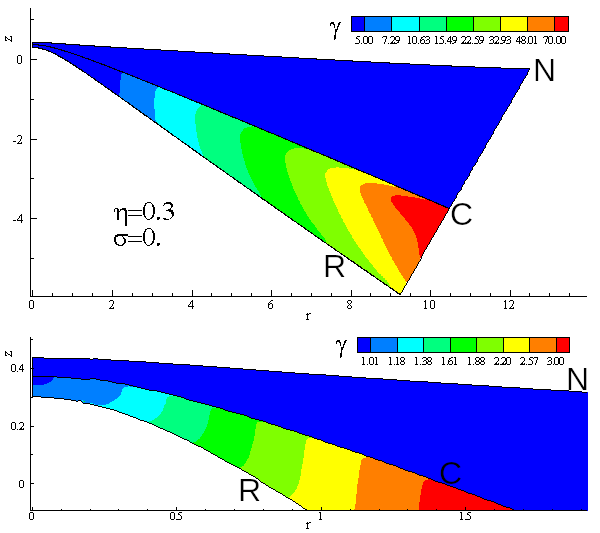}
 \caption{Shocked flow of plasma from the companion star(in the cone NC) and the pulsar (in the cone RC) for $\sigma=0$ and $\eta=0.3$. The radial flows of the supersonic plasma from the companion star (located at the point (0,1)) and the pulsar (located at the point (0,0)) upstream of the shocks are not shown. The flows of the  non-relativistic and relativistic plasma in the post shock region are separated by the \ac{cd} (line C). The color in the region of the flow of the shocked relativistic plasma shows the Lorentz factor of the plasma. {Panels show a large-scale structure and a zoomed view of the inner part of the binary system. }}
 \label{fig1}
\end{center}
\end{figure}

\begin{figure} 
\begin{center}
\includegraphics[width=100mm]{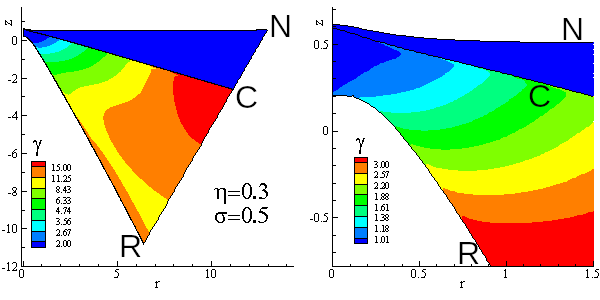}
 \caption{Same as in Fig.\ref{fig1} for $\sigma=0.5$.}
 \label{fig2}
\end{center}
\end{figure}

\begin{figure} 
\begin{center}
\includegraphics[width=100mm]{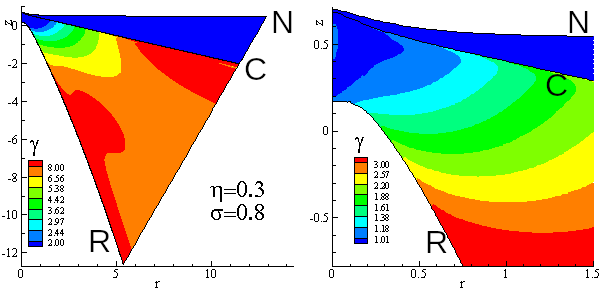}
 \caption{Same as in Fig.\ref{fig1} for $\sigma=0.8$.}
 \label{fig3}
\end{center}
\end{figure}

Numerical simulations of the flow geometry and the bulk Lorentz factor are shown in Figs.~\ref{fig1},~\ref{fig2}, and
\ref{fig3} for \(\sigma=0\), \(0.5\), and \(0.8\), respectively (\(\eta=0.3\) was fixed for all three cases). {The spatial coordinates are dimensionless in the star separation units.} The
optical star and the pulsar are located at the points with coordinates \((r=0,z=1)\) and \((r=0,z=0)\),
respectively. Qualitatively the flow structure is identical to those revealed with other
simulations \citep{bakuk08,bakuk12,2012A&A...544A..59B,2015A&A...577A..89B}. Supersonic winds propagate from the stars until
reaching bow-shaped \acp{ts}, which are marked in the figures with symbols ``R'' and ``N'' (relativistic pulsar wind
\ac{ts} and non-relativistic stellar wind \ac{ts}, respectively).  The supersonic flows are not shown in the figures.
The volume between the \acp{ts} is filled with shocked gas. A \ac{cd}, which is marked with symbol ``C'' in the figures, separates the relativistic and the non-relativistic gas.
The shocked nonrelativistic plasma originated from the companion star propagates in the cone ``NC,'' and the shocked relativistic plasma in the cone ``CR.''

From a comparison of the figures it can be seen that the increasing wind magnetization has a considerable impact on the
flow geometry.  The region occupied by the shocked pulsar wind expands significantly. This phenomena is accompanied by a
decrease of the bulk Lorentz factor. After passage of the \ac{ts}, a fraction of the kinetic energy
is transformed into Poynting flux that results in a 
decrease of the Lorentz factor of the shocked wind.

\begin{figure} 
\begin{center}
\includegraphics[width=100mm]{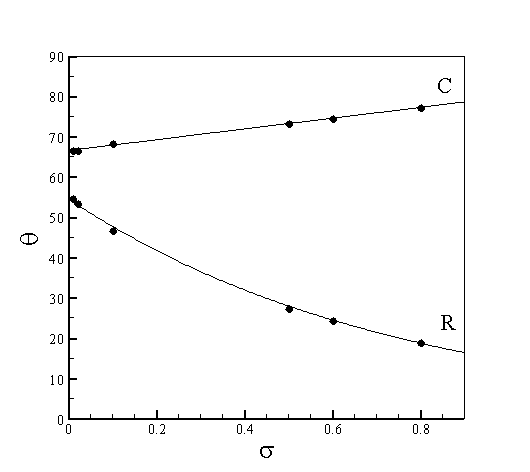}
 \caption{Dependence of the opening angle of the \ac{cd} (marked with ``C'') and the relativistic \ac{ts} front (marked with ``R'') on $\sigma$.}
 \label{fig4}
\end{center}
\end{figure}

In Fig.~\ref{fig4}, we show the dependence of the opening angles of the relativistic \ac{ts} and the \ac{cd} on $\sigma$. The
opening angle of the \ac{cd} increases and the opening angle of the relativistic \ac{ts} decreases with $\sigma$. This
results in a significant expansion of the region occupied by the shocked winds.
According to Eq.~(\ref{energy}), the  increase of $\sigma$ results into growth 
of the actual total pulsar energy losses. The flux of the kinetic energy remains constant. If there were no impact from the magnetic collimation, one would expect that the impact on the flow structure can be approximately determined by the effective change \(\eta\)-parameter:
\begin{equation}
  \eta_{\rm eff}=\eta\left(1+\frac23\sigma\right)\,.
\end{equation}
In the regime \(\eta_{\rm eff}>10^{-2}\), the opening angle of the \ac{cd} can be approximately described \citep[see {Fig.11} in][]{bakuk08} as:
\begin{equation}
  \theta_\cd\approx85^\circ + 15^\circ\ln\eta_{\rm eff}\approx67^\circ + 10^\circ\sigma\,,
\end{equation}
where we used \(\eta=0.3\).
Thus, the increase of the opening angle of the \ac{cd} seen in Fig.~\ref{fig4} is consistent with the change of the
pulsar spindown power due to the contribution from Poynting flux. Based on result from \citet{bakuk08}, one should also expect an increase of the pulsar wind
\ac{ts} as the \(\sigma\)-parameter grows. In contrast, as seen in Fig.~\ref{fig4} the opening angle of the pulsar
wind \ac{ts} is decreasing.

\section{Discussion}

\begin{figure} 
\begin{center}
\includegraphics[width=80mm]{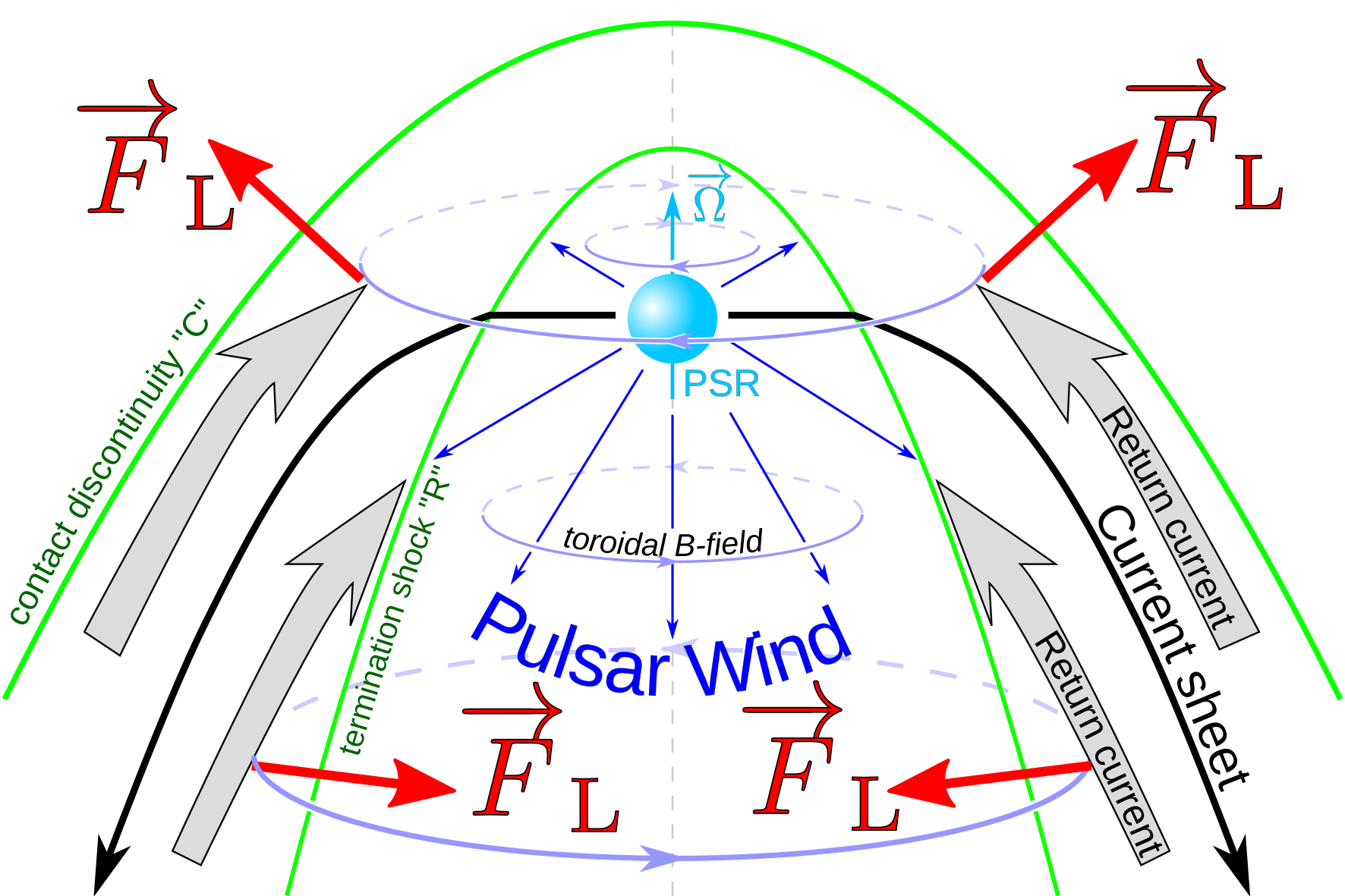}
 \caption{Scheme of the flow of the relativistic plasma. The Lorentz force, \(\vec{F}_{\rm L}\),   extends the cone of the contact discontinuity (C) and collimates the relativistic shock wave (R). }
 \label{fig5}
\end{center}
\end{figure}

Numerical simulations of the interaction of the magnetized pulsar wind with the isotropic stellar wind under the
approximation of axisymmetry show that the wind magnetization has a strong impact on the interaction geometry in the
regime \(\sigma\sim1\). As shown in Fig.~\ref{fig4}, the solid angle occupied by the shocked pulsar wind tends to
increase considerably with increasing \(\sigma\).  The reason of this dependence can be qualitatively understood as the
Lorentz force impact \citep[see, e.g.,][]{2012MNRAS.420.2793K}. In Fig.~\ref{fig5}, we show a scheme showing the
structure of the relativistic outflow. For the sake of simplicity, we assumed that the pulsar magnetic axis coincides
with the rotation axis. The magnetic and rotational axes of the pulsar are aligned along the pulsar-star connecting line.  In
the unshocked wind zone, the current sheet, i.e., the surface that separates regions of opposite polarity 
toroidal magnetic field (shown with a solid black line in Fig.~\ref{fig5}), lies on the equatorial plane, which is
perpendicular to the symmetry axis. The current sheet propagates through the \ac{ts}, but gets deflected in the shocked relativistic
plasma.  The electric currents propagate through the shocked pulsar wind above and below the current sheet as shown in
Fig.~\ref{fig5}. Since the polarity of the magnetic field below and above the current sheet changes, the Lorentz
force is directed differently.
Therefore, the collimating force above the current sheet is directed towards the \ac{cd} and creates an additional pressure
on it. This contributes to the growth of the \ac{cd} opening angle. The collimating force below the current sheet
is directed towards the pulsar wind \ac{ts} front. Consequently, the \ac{ts} is pushed  closer to the symmetry axis.
 
Although we considered a rather specific case, in which the rotation and magnetic axes of the pulsar are aligned with
the symmetry axis, a similar argument seems to be valid in a more general case. At large distances, pulsar winds practically
do not differ from the winds launched by a split monopole magnetic field  \citep{2012MNRAS.420.2793K}.  The  distribution 
of  magnetic pressure in the flow from a split
monopole does not depend on the orientation of the magnetic axis \citep{1973ApJ...180..207M,1999A&A...349.1017B}. Thus, the structure
of the Lorentz force should remain unchanged independently on the orientation of the pulsar magnetic axis.

The influence of the orientation of the rotation axis needs to be studied in the \ac{3d} approximation \citep[see, e.g.,
the study by][for the case of bow-shock \ac{pwn}]{2019MNRAS.484.4760B,2019MNRAS.485.2041B,2019MNRAS.484.5755O}.  To consider such a configuration in a binary system,
additional studies are required, which are, however, beyond the scope of this paper.

\ac{grbs} represent a class of gamma-ray sources where \ac{mhd} processes are expected to play an essential
role. Broadband \ac{nt} components observed from these systems are most likely produced in the shocked pulsar
wind. Thus, one expects important similarities between \acp{grbs} and \acp{pwn} \citep[see,
e.g.,][]{1994ApJ...433L..37T,1999APh....10...31K,2007MNRAS.380..320K}. There are, however, also important differences
that include a significantly smaller scale, the presence of dense photon target field, and gradual changes of the
physical conditions due to orbital motion. The high pressure as well as the dense photon target field reduce the
relevant space- and time-scales by many orders of magnitude, allowing one to study the processes taking place in
\acp{pwn} under varying conditions and on much shorter time scales. \ac{grbs} can be considered as essential physical
laboratories to study the physics of \ac{pwn}.

The strong orbital phase dependence of the emission is expected to be caused by several factors. In particular this
includes the following: (i) anisotropic \ac{ic} scattering of target photons provided by the optical companion
\citep{2008MNRAS.383..467K,2008A&A...477..691D}, (ii) changing rate of adiabatic and radiative losses
\citep{1999APh....10...31K,2007MNRAS.380..320K}, (iii) orbital phase depended gamma-gamma attenuation
\citep{2005ApJ...634L..81B,2006A&A...451....9D}, (iv) Doppler boosting caused by bulk re-acceleration of the shocked
pulsar wind \citep{2012ApJ...753..127K,mitya14}. The latter factor is of special interest, as this is the only factor
that can enhance the source apparent luminosity above the limit determined by the spindown luminosity. Importantly,
observations of \ac{psrb} with \ac{fer} have shown that the gamma-ray flux level may increase above this limit
\citep[see the discussion in][]{2015ApJ...811...68C,2018ApJ...863...27J,2018MNRAS.479.4849M}. Such a detection makes
models that do not involve Doppler boosting less feasible
\citep[e.g.,][]{2012ApJ...752L..17K,2013A&A...557A.127D}. However, in the context of \ac{grbs}, the Doppler boosting
affects the synchrotron and \ac{ic} components in a very different way \citep[see, e.g.,][]{mitya14}. The weakening of
the magnetic field, which is expected in accelerating flow, suppresses the synchrotron emission. Thus, Doppler boosting
can considerably enhance only the \ac{ic} emission. Note however that, since the dominant photon target is provided by
the optical companion, the enhancement factor is different compared to those obtained for blob sources \citep[see,
e.g.,][]{2018MNRAS.481.1455K}.

 Doppler boosting in \ac{grbs} critically depends on the solid angle into which the shocked pulsar wind
propagates. \citet{bakuk08} obtained analytic approximations that allows us to describe the flow geometry in the case of
weak magnetization of the pulsar wind. Significant collimation, \(4\upi/\Delta\Omega>10\), can be achieved only in the cases of small \(\eta\)
parameters (\(\eta<0.05\)), which implies an unrealistically powerful stellar wind. In this paper, we studied the impact of a high wind
magnetization on the collimation of the shocked pulsar wind. Our simulation indicates that higher values of the
\(\sigma\)-parameter, \(\sigma\sim1\), even result in further de-collimation of the shocked pulsar wind. This finding is consistent
with the conventional theories of the dynamics of magnetized outflows. We therefore conclude that the
de-collimation is caused by physical reasons and it is unlikely to be a numerical artifact. This challenges the
scenarios that adopt Doppler boosting as the main mechanism responsible for production of GeV flares detected with
\ac{fer} from \ac{psrb}.

In this paper we performed numerical simulations of the collision of a magnetized anisotropic pulsar wind with the
circumbinary environment under the approximation of an axisymmetric ideal \ac{mhd} flow. The magnetic field in the pulsar
wind was assumed to be torroidal, with strength depending on the distance to the pulsar, \(r\), and the polar angle,
\(\theta\): \(B_\varphi\propto \sin\theta/r\). The simulations were performed for three equatorial magnetizations of the
pulsar wind: \(\sigma=0.5\), \(0.6\), and \(0.8\), and extend the study by \citet{bakuk12} for the range of
\(\sigma\leq0.1\).  While in the case of small wind magnetization, \(\sigma\leq0.1\), the magnetic field has a minor
impact on the plasma dynamics \citep{bakuk12}, in the newly considered regime the influence of the magnetic field
appears to be considerable. The opening cone in which the shocked pulsar wind propagates increases with growth of the
wind magnetization. This effect can be explained as the impact of the Lorentz force that collimates the part of the flow
below the current sheet and expands the part that is above (see Fig.~\ref{fig5}). This finding has an important
implication for scenarios that aim to explain the origin of the bright GeV flares detected with \ac{fer} from
\ac{psrb}. Since the registered gamma-ray luminosity exceeds the pulsar spindown luminosity, the production of the
gamma-ray emission in highly collimated outflows seems to be the most plausible scenario. Our simulations show that in
the considered regime, \(\sigma\sim1\), the magnetic pinch results in de-collimation of the outflow. This makes
scenarios relaying on a large Doppler boosting to be less feasible and the overall interpretation of the GeV flares
to be even more challenging.

\section*{Acknowledgments}
The authors thank V.Bosch-Ramon, M.V.Barkov, and referee, Barbara Olmi, for their useful comments.
The work of S.Bogovalov was supported by the Ministry of Education and Science of the Russian Federation, MEPhI Academic Excellence Project (contract \textnumero 02.a03.21.0005, 27.08.2013) and by RFBR  grant \textnumero 16-02-00822/18.
DK is supported by JSPS KAKENHI Grant Numbers JP18H03722, JP24105007, and JP16H02170. AVK acknowledges support by RFBR 18-02-00907.

\label{lastpage}

\end{document}